\documentclass[aps,prb,twocolumn,superscriptaddress,floatfix,longbibliography]{revtex4-1}
\usepackage{amsfonts}
\usepackage{amssymb}
\usepackage{graphicx}
\usepackage{dcolumn}
\usepackage{bm}
\usepackage{amsmath}
\usepackage[colorlinks,linkcolor=magenta,citecolor=blue,urlcolor=blue]{hyperref}
\usepackage{changes}
\usepackage{float}

\begin{document}

\title{Types of dynamical behavior in a quasiperiodic mosaic lattice}
\author{Yu Zhang}
\affiliation{Beijing National Laboratory for Condensed Matter Physics, Institute
of Physics, Chinese Academy of Sciences, Beijing 100190, China}
\affiliation{School of Physical Sciences, University of Chinese Academy of Sciences,
Beijing, 100049, China}
\author{Chenguang Liang}
\affiliation{Beijing National Laboratory for Condensed Matter Physics, Institute
of Physics, Chinese Academy of Sciences, Beijing 100190, China}
\affiliation{School of Physical Sciences, University of Chinese Academy of Sciences,
Beijing, 100049, China}
\author{Shu Chen}
\email{schen@iphy.ac.cn}
\affiliation{Beijing National Laboratory for Condensed Matter Physics, Institute of
Physics, Chinese Academy of Sciences, Beijing 100190, China}
\affiliation{School of Physical Sciences, University of Chinese Academy of Sciences,
Beijing, 100049, China}
\date{\today }

\begin{abstract}
Quasiperiodic mosaic systems with the quasiperiodic potential being added periodically with a fixed lattice interval have attracted significant attention due to their special spectral properties with exactly known mobility edges, which separate localized and extended states. These mobility edges do not vanish even in the region of large quasiperiodic potential strength,  although  the width of the energy window of extended states decreases with the increase in potential strength and thus becomes very narrow in the limit of strong quasiperiodic disorder.
In this paper, we study the dynamics of a quasiperiodic  mosaic lattice and unravel its peculiar dynamical properties. By scrutinizing  the expansion dynamics of wave packet and the evolution of density distribution, we unveil that the long-time density distribution  displays obviously different behaviors at odd and even sites in the region of large quasiperiodic potential strength. Particularly, the time scale of dynamics exhibits an inverse relationship with the quasiperiodic potential strength.
To understand these behaviors,  we derive an effective Hamiltonian in the  large quasiperiodic potential strength region, which is composed of decoupled Hamiltonians defined on the odd and even sites, respectively. While all eigenstates of the effective Hamiltonian defined on even sites  are localized,  the eigenstates of effective Hamiltonian defined on odd sites include both localized and extended eigenstates. Our results \textcolor{red}{suggest} that the effective Hamiltonian can describe the dynamical behaviors well in the  large quasiperiodic potential strength region and provides an intuitive framework for understanding the peculiar dynamical behaviors in the quasiperiodic mosaic lattice.
\end{abstract}

\maketitle

\section{Introduction}

Quasiperiodic systems, as pivotal platforms for investigating Anderson localization \cite{Anderson,Anderson50,Mirlin,Lee,2D} and mobility edges \cite{MEAnderson1,MEAnderson2,Mott1}, have attracted considerable attention and extensive exploration.
Compared to the Anderson model in which the extended-localized transition and mobility edge only exist in three dimension \cite{2D, MEAnderson1,MEAnderson2,Mott1, Experiment1, Experiment4,Experiment5}, quasiperiodic systems can host the the extended-localized transition and mobility edge in one dimension \cite{AAmodel,AAHmodel,Cai,HYao,MobilityEdge1, self-consisitent,Experiment2, Experiment3,XCXie,XCXie1,SarmaMobilityEdge1,SarmaMobilityEdge2,GaoXL}.
While the Aubry-Audr\'{e} (AA) model hosts extended-localized transition \cite{AAmodel,AAHmodel} at the self-dual point, generalizations of the AA model \cite{MobilityEdge1, self-consisitent,XCXie,XCXie1,SarmaMobilityEdge1,SarmaMobilityEdge2,Deng, longrange,MosaicAA, globaltheory1, hidden-duality, RG, Critical, DDVu1,LiuXJ,WangLi} or periodically driving the quasiperiodic systems \cite{Qin,Driving, Driving1, Driving2, Driving3, Driving4} can give rise to fruitful phenomena of mobility edges. Recently,
a class of quasiperiodic mosaic models \cite{MosaicAA}, in which the quasiperiodic potential periodically occurs with a fixed interval, was proposed as another type of quasiperiodic systems with exact mobility edges.
Different from previous models \cite{SarmaMobilityEdge1,SarmaMobilityEdge2}, the quasiperiodic mosaic models have mobility edges even in the  large quasiperiodic potential strength region and can host multiple mobility edges. Stimulated by this quasiperiodic mosaic model, many other mosaic systems have been studied and show quite interesting behaviors \cite{LiuYX,Longhi,MosaicSystem1,MosaicSystem2,MosaicSystem3,MosaicSystem4,MosaicSystem5,Mosaic6,Dwiputra}. Recently, the experimental realization of the  quasiperiodic  mosaic model was also reported \cite{mosaicAAExperiment}.

While the diffusion of a wave packet in a uniform lattice is ballistic, the Anderson localization can suppress the diffusion of the wave packet.  For a one-dimensional quasiperiodic lattice with mobility edge, for example, the Ganeshan-Pixley-Das Sarma (GPD) model \cite{SarmaMobilityEdge2}, there exists an intermediate regime in which the extended and localized states coexist and are separated by the mobility edges. The dynamical behaviors in this region are quite subtle and have drawn considerable attention, showcasing unique phenomena \cite{Dynamics1, Dynamics2, Dynamics3}. It was shown that the  dynamical behaviors of a wave packet in the intermediate regime of the GPD model exhibit a blend of both localized and extended features \cite{Dynamics1}, distinguishing them from dynamics in the multifractal region.
For  quasiperiodic  mosaic systems, despite Avila's global theory \cite{GlobalTheory}providing an exact expression for the mobility edge, the dynamics of the  quasiperiodic  mosaic model has not yet been studied and is worth further exploration.

In this paper, we investigate the dynamics of a quasiperiodic mosaic model across a range of quasiperiodic potential strengths and particularly unravel the peculiar dynamical behavior in the large quasiperiodic potential strength limit. Our analysis involves visually tracking the evolution of a Gaussian wave packet and study the evolution of particle number distribution with different initial states. We observe quite distinct behaviors under different quasiperiodic potential strengths.  Particularly,  in the large quasiperiodic potential strength limit, we observe significant density distribution differences between even and odd sites, with the expansion timescale of the wave packet showing a linear relationship with the quasiperiodic potential strength $\lambda$.
To understand the physical original of these peculiar dynamical behaviors, we derive an effective Hamiltonian in the large quasiperiodic potential strength region, indicating that the effective Hamiltonian for the odd and even sites are decoupled.
This effective Hamiltonian provides a clear physical picture of the even-odd difference of the dynamical behaviors and effectively captures the scaling behavior of eigenvalues, as well as the property of extended-localized transition. Additionally, analysis of the eigenstates of the effective Hamiltonian yields further insights into the understanding of the spectral structure of the quasiperiodic mosaic model. 
Our study sheds light on the physical properties of the quasiperiodic mosaic model, unraveling the physical origin of the peculiar dynamical behaviors in the quasiperiodic mosaic model. 

The rest of the paper is organized as follows: In Sec.\ref{Model}, we introduce the quasiperiodic mosaic model and its main properties briefly.
Then, we study the dynamics of the wave packet expansion and the evolution of particle number distribution with different initial states.
In Sec.\ref{Heff}, we derive the effective Hamiltonian in the large quasiperiodic potential strength limit and analyze the properties of the effective Hamiltonian, through which we can understand clearly the even-odd difference of the wave packet dynamics. Finally, in Sec. \ref{E}, we give a summary of our findings. 

\section{Model and results \label{Model}}

\subsection{Model }
We consider the quasiperiodic mosaic model described by the following Hamiltonian:
\begin{align}
& \hat{H}=J \sum_j(\hat{c}^\dagger_j \hat{c}_{j+1}+H.c.) + 2 \sum_j\lambda_j \hat{n}_j, \\
& \lambda_j= \{\begin{array}{lc}
       \lambda \cos(2\pi \omega j+ \phi) , & \mod(j,2)=0,\\
      0, &\text{otherwise},
\end{array}
\end{align}
where $\hat{c}_j^\dagger (\hat{c}_j)$ is the fermion creation(annihilation) operator and $\hat{n}_j=\hat{c}^\dagger_j \hat{c}_j$ is the particle number operator at the $j$th site.
$\lambda$ is the strength of  quasiperiodic potential and $J$ is the hopping amplitude, for which we set $J=1$ as the unit of energy in the following calculation.
$\omega$ is an irrational number and we choose $\omega=\frac{\sqrt{5}-1}{2}$.
$\phi$ is a random phase which leads to the realization of different quasiperiodic potentials.
We note that the quasiperiodic potential is distributed only on even sites.

\begin{figure*}
    \centering
    \includegraphics[width=1\textwidth,height=0.5\textwidth]{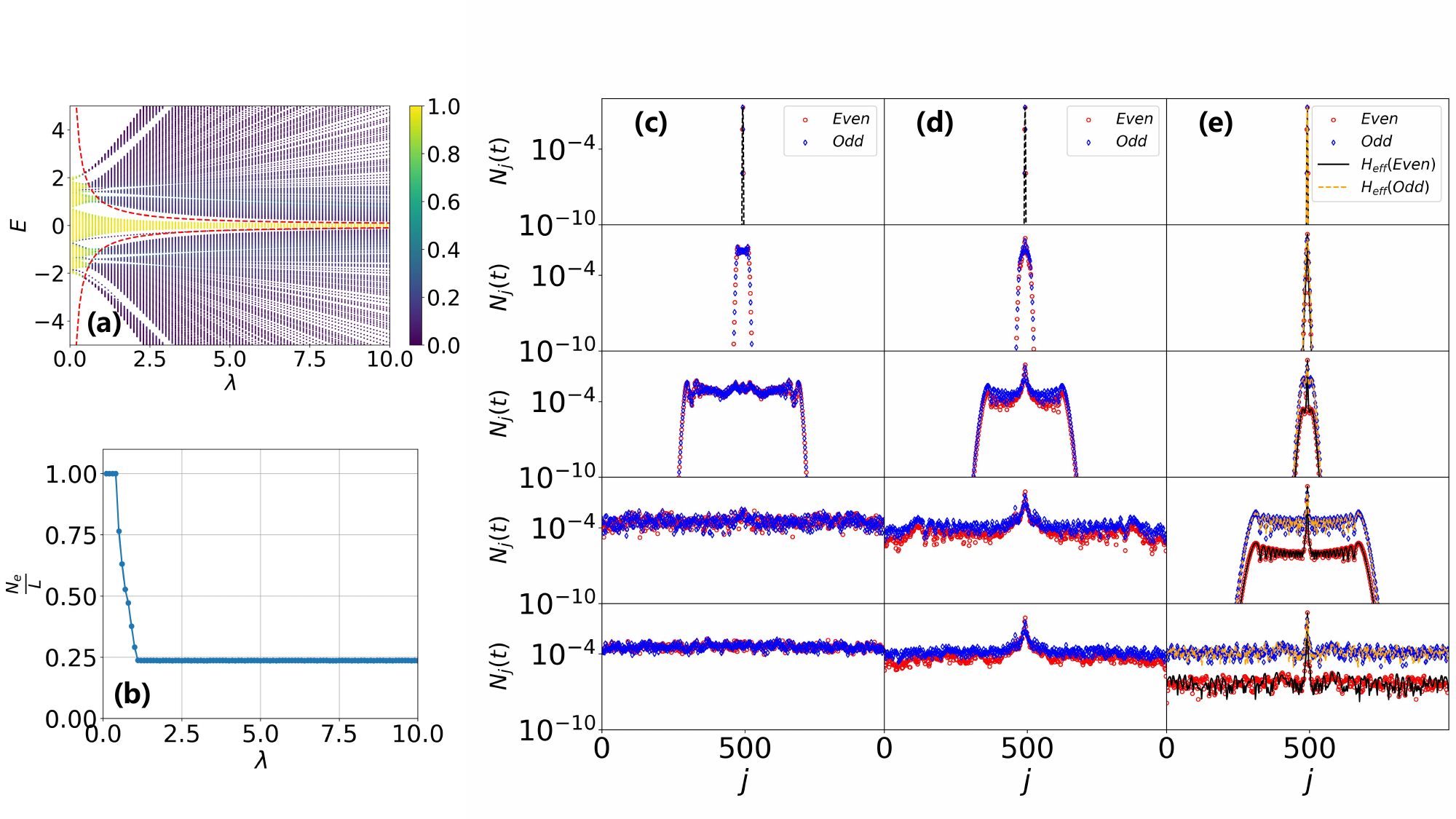}
    \caption{Basic properties and wave packet dynamics of the quasiperiodic mosaic model. (a) The eigenvalue spectrum versus $\lambda$ with the fractal dimension of corresponding eigenstate denoted by $D_2$. (b) The proportion of extended states  in quasiperiodic mosaic model.
    (c)-(e) The evolution of wave packet under different quasiperiodic potential strength. From left to right, we choose $\lambda=0.1,1,10$. From top to bottom, we choose $t=0,10,10^2,10^3,10^4$.
    The red circle (blue diamond) markers represent the even (odd) wavefunction distribution. We choose the Gaussian wave packet with $\sigma=1$ as the initial state.
    In (e), the solid(dashed) lines represent the even (odd) wavefunction distribution from the evolution of the effective Hamiltonian.
    We choose $L=987$ in our calculation and take the average over 100 realizations of the quasiperiodic potential.  }
    \label{figure1}
\end{figure*}

The exact mobility edges for the quasiperiodic  mosaic model have been determined by applying Avila's global theory \cite{MosaicAA} and the expression of the exact mobility edges is  $E_c=\pm \frac{1}{\lambda}$. 
To get a straightforward view, we display the eigenvalues and the fractal dimension of the corresponding eigenstates of the quasiperiodic mosaic model in Fig.\ref{figure1}(a) with the dashed lines  indicating the mobility edge predicted by the Avila's global theory. The fractal dimension is defined  using the inverse participation ratio (IPR)
\begin{equation}
I_{2}(n)=\sum_j |\psi_{n,j}|^{4} \propto L^{-D_2},
\label{moment}
\end{equation}
where $\psi_{n,j}$ represents the $n$th eigenstate's amplitude on the $j$th site.
The fractal dimension $D_2$ takes different values at the thermodynamic limit in various regions: while $D_2 \rightarrow 1$ in the extended region, $D_2 \rightarrow 0$ in the localized region. 

It is shown that the energy window of extended states becomes narrower with the increase in the strength of  quasiperiodic potential. Although the energy window is very narrow in the strong limit of $\lambda$, extended eigenstates always exist in the large quasiperiodic potential strength region. To see it clearly, we show the proportion of extended states $\frac{N_e}{L}$ with different quasiperiodic potential strengths in Fig.\ref{figure1}(b), where $N_e$ is the number of extended eigenstates. If $\frac{N_e}{L}=1 ~(0)$, all the eigenstates  are extended (localized).
When the quasiperiodic potential strength $\lambda$ increases to a certain value, the proportion of extended states no longer changes with the quasiperiodic potential strength increases.


\subsection{Wave packet dynamics}

To see how the change of spectrum and mobility edges affect the dynamical evolution, we shall study the wave packet dynamics in the mosaic lattice.
We choose the Gaussian wave packet at the center of the system as the initial state, which is described by
\begin{align}
|\psi(0) \rangle= C \sum_j \exp(-(j-j_c)^2/ 2\sigma^2) |j\rangle,
\end{align}
where $\sigma$ is related to the width of the Gaussian wave packet, $j_c=\frac{L}{2}$ is the center of the system size,
$| j \rangle = \hat{c}_j^\dagger |0\rangle $ and $|0\rangle$ is vacuum state.
$C$ is a normalization factor to ensure $\langle \psi(0)|\psi(0) \rangle=1$ .
The  time evolution of the wave packet is determined by the Hamiltonian and the wave function at time $t$ can be written as:
\begin{align}
|\psi(t) \rangle=e^{-i \hat{H} t}|\psi(0) \rangle.
\end{align}

To get an intuitive picture of the evolution of wave packets for various $\lambda$, we show the distribution of particle number at different times $N_j(t)=\langle \psi(t)| \hat{n}_j| \psi(t) \rangle$.
The results are shown in Fig.\ref{figure1}(c-e), which exhibit the wave packet dynamics across varying degrees of quasiperiodic potential strength.
In Fig. \ref{figure1} (c), with $\lambda=0.1$, all the eigenstates are extended. The wave packet expands and lose all the initial information after a long time evolution. Here we adopted logarithmic coordinates for $N_j(t)$. Notably, there are no discernible differences between distributions at even and odd sites.
In Fig.\ref{figure1} (d), we choose $\lambda=1$ , which is located in the region with mobility edges and the number of extended states decreasing with the increase of the quasiperiodic potential strength. We observe a slight difference between distributions at odd and even sites in the long time limit.
Finally, in Fig. \ref{figure1}(e), we have $\lambda=10$, which is in the region with the number of extended states keeping invariant with the increase of $\lambda$. Here, a clear discrepancy between distributions at even and odd sites emerges, maintaining throughout evolution while retaining identical propagation velocities for both sites.
Additionally, in Figs. \ref{figure1}(c)-\ref{figure1}(e), the wave packet gradually expands over time. After a sufficiently long evolution time, a portion of the wave packets in the center remains localized at their initial positions, while another portion spreads across the entire system, attributed to the presence of localized eigenstates,  as shown in Figs. \ref{figure1}(d)-\ref{figure1}(e). It is noteworthy that while wave packet expansion occurs in all cases, the time scales of expansion are different.

\begin{figure}
    \centering
    \includegraphics[width=0.45\textwidth,height=0.35\textwidth]{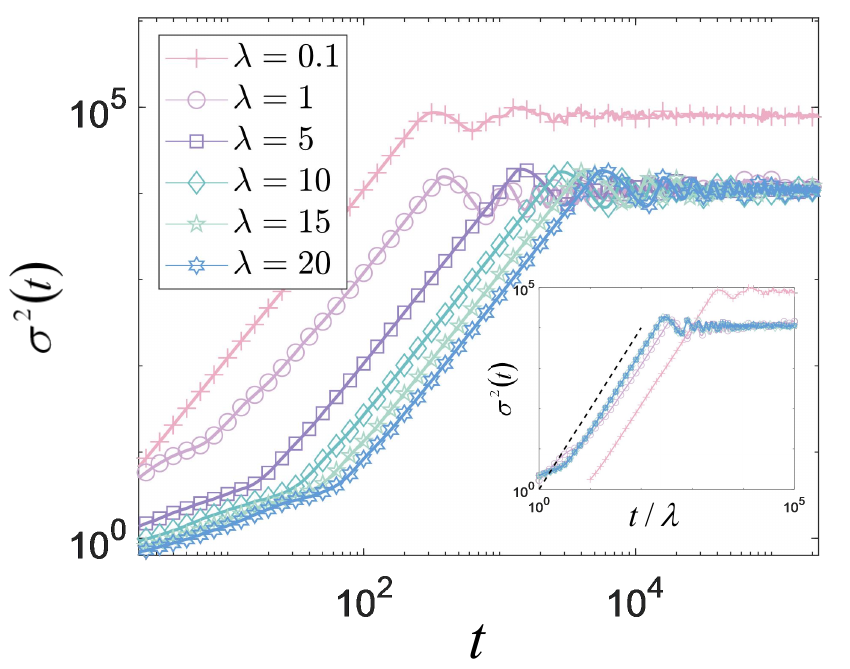}
    \caption{ Growth of $\sigma^2(t)$ under different quasiperiodic potential strength.
    We choose different quasiperiodic potential strength from $\lambda=0.1$ to $\lambda=20$, $L=987$ and $\sigma=1$.
    We take the average over 100 realizations of the quasiperiodic potential in our calculation.
    In the inset, we have chosen a rescaled $x$-axis ($t/\lambda$). It is shown that all lines in the large quasiperiodic potential strength region coincide,
    suggesting that the time scale of the dynamic evolution is proportional to $\lambda$. The dashed line is $\sigma^2(t) \propto t^2$.
    }
    \label{X2}
\end{figure}

To quantitatively describe the expansion of wave packets and the time scales of the dynamics, we consider the mean-square displacement \cite{Abe,Piechon,TongPQ}, which is defined as
\begin{align}
\sigma^2(t)=\langle \psi(t)|(\hat{X}-j_c)^2| \psi(t) \rangle,
\end{align}
where $\hat{X}$ is the coordinate operator.
In Fig.\ref{X2}, we display the growth of $ \sigma^2(t) $ under different strengths of quasiperiodic potential.
The system with larger quasiperiodic potential strength will expand more slowly: to achieve the same wave packet width, it requires longer time for larger $\lambda$.
Although having different time scales,  the dynamical behaviors in the large quasiperiodic potential strength region are quite similar.
In the inset of Fig.\ref{X2}, we find the lines in the large quasiperiodic potential strength region coincide when we choose a rescaled $x$-axis ($t/\lambda$).
That means the time scales of wave packet dynamics are proportional to the quasiperiodic potential strength $\lambda$.
In other words, the quasiperiodic strength can linearly modulate the time scale of wave packet dynamics.
After a short increase, we find that the linear interval in Fig.\ref{X2} can be well fitted by $\sigma^2(t) \propto t^2 $ before they reach saturation values, which indicates the transport of the wave packet are ballistic all for different  quasiperiodic potential strengths.

\subsection{Particle number distribution}

Next we study the time evolution of particle number distribution for a half-filled system.
By choosing different initial states and counting the number of particles at different sites, we can calculate the two-point correlation function $\hat{\rho}_{ij}=\langle  \psi(t)|\hat{c}^\dagger_i \hat{c}_j|\psi(t)\rangle$.
The details of the calculation of the two-point correlation function can be found in Appendix \ref{Imbalance}. The diagonal part of $\hat{\rho}_{ii}$ gives particle number distribution $\hat{n}_i$.

The first quantity we considered is the imbalance \cite{ImbalanceA}, which is defined as:
\begin{equation}
\mathcal{I}(t)=\frac{N_{\text{even}}(t)-N_{\text{odd}}(t)}{N_{\text{even}}(t)+N_{\text{odd}}(t)},
\end{equation}
where
\begin{align}
  N_{\text{even(odd)}}(t)=\sum_{j=\text{even(odd)}}\langle\psi(t)|\hat{n}_j|\psi(t) \rangle
\end{align}
counts particle number distributed on the even (odd) sites at time $t$.
Imbalance can naturally describe the difference in particle number distribution of odd and even sites.
\begin{figure}
    \centering
    \includegraphics[width=0.5\textwidth,height=0.4\textwidth]{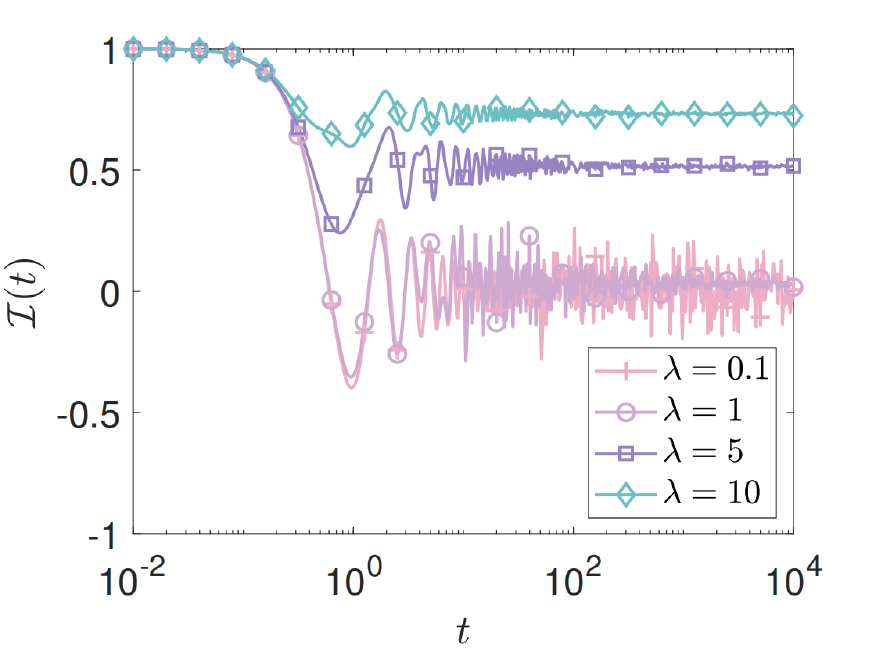}
    \caption{The evolution of imbalance $\mathcal{I}(t)$ for different quasiperiodic potential strength. We choose $L=100$  in our calculation and take the average over 100 realizations of the quasiperiodic potential.
    Besides, we place different markers on the lines with different $\lambda$ to differentiate them.
     }
    \label{ImbalanceFig}
\end{figure}
Choose the initial state as the product state given by
\begin{equation}
|\psi(0)\rangle =\prod_{j=1}^{L/2} \hat{c}^\dagger_{2j-1} |0\rangle, \label{initial1}
\end{equation}
corresponding to $\mathcal{I}(0)=1$ with the odd sites being fully occupied and the even sites unoccupied.
We show the evolution of imbalance with various quasiperiodic potential strengths $\lambda= 0.1,1, 5$ and $10$ in Fig.\ref{ImbalanceFig}.
In the weak and intermediate region ($\lambda=0.1$ and $1$), $\mathcal{I}(t)\approx 0$ after a long time evolution. When $\lambda$ is located in the large quasiperiodic potential strength  region, the Imbalance obviously deviates zero. With the increase of the quasiperiodic potential strength, the value of imbalance in the long time becomes larger, consistent with the results in wave packet dynamics. By detecting this quantity, we can clearly see the difference in particle number distribution at odd and even sites.

To detect the effect of the quasiperiodic potential strength on the time scale of the system,  we consider another initial state given by
\begin{equation}
|\psi(0)\rangle =\prod_{j=1}^{L/2} \hat{c}^\dagger_{j} |0\rangle   \label{initial2}
\end{equation}
and  calculate the value $N_h(t)$, reflecting how many particles remaining in the left half of the system at time $t$:
\begin{align}
N_h(t)=2\sum_{j=1}^{L/2} \langle \psi(t)|\hat{n}_j |\psi(t)\rangle/ L.
\end{align}

\begin{figure}
    \centering
    \includegraphics[width=0.5\textwidth,height=0.4\textwidth]{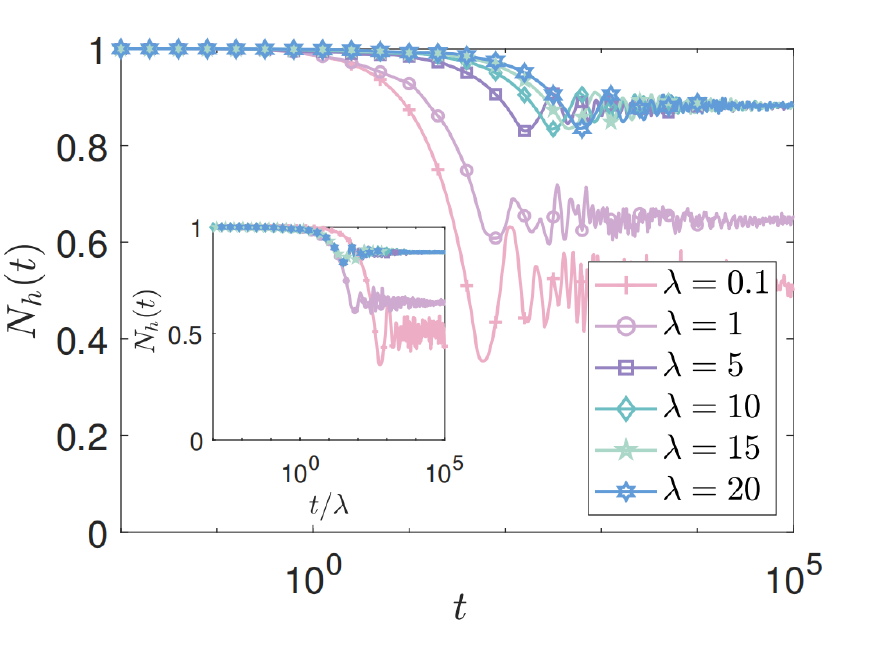}
    \caption{Time evolution of $N_h(t)$ for different disorder strength. We choose $L=100$ in our calculation and take the average over 100 realizations of the quasiperiodic potential.   }
    \label{Imbalance1}
\end{figure}

While $N_h(0)=1$ for the initial state,  $N_h(t)$ decreases with the time evolution and approaches to a stable value in the long time limit, as illustrated in Fig.\ref{Imbalance1}.
In the large quasiperiodic potential strength region, the saturation values are the same for different quasiperiodic potential strengths.  This is because the underlying structures (fractions of localized or extended eigenstates) remain the same and independent of
$\lambda$ in this region.
However, the time scale changes with the increase of the quasiperiodic potential strength.
When we rescale the x-axis to $t/\lambda$, we find these curves coincide in the large quasiperiodic potential strength region.


\section{Effective Hamiltonian in the large $\lambda$ limit\label{Heff}}
In the large quasiperiodic potential strength region, we have unveiled that the dynamics in the quasiperiodic mosaic lattice exhibits some special  properties, including differences in density distribution between even and odd sites as well as time scales linearly depend on the strength of quasiperiodic potential.
To understand these special dynamical behaviors, we shall provide an explanation by using an effective Hamiltonian in the subsequent discussion.
Suppose that the quasiperiodic potential is sufficiently large, surpassing the strength of the hopping terms. We can apply the Schrieffer-Wolf (S-W) transformation method to construct the effective Hamiltonian for the quasiperiodic mosaic model. The effective Hamiltonian can provide a clear physical insight for understanding  the special dynamical behaviors in the large quasiperiodic potential strength region.

\subsection{Derivation of the effective Hamiltonian}

The S-W transformation decouples two weakly coupled subspaces through a unitary transformation \cite{SW, SW1, SW2}.
For the case with large quasiperiodic potential strength, the hopping term is small in comparison with the onsite quasiperiodic potential.
Thus we can decouple the Hamiltonian into two decoupled chains by applying the S-W transformation.
The Hamiltonian can be written as:
\begin{align}
    \hat{H}= \textbf{c}^\dagger H \textbf{c} ,
\end{align}
where $\textbf{c} = (\hat{c}_1, \hat{c}_2,\dots \hat{c}_L)^T $ and $H$ is the matrix form of the Hamiltonian.
We consider an anti-Hermitian matrix with the matrix elements given by
\begin{align}
S_{mn}=&\sum_p [\frac{J}{\lambda_p}(\delta_{m,p}\delta_{n,p+1}+\delta_{m,p}\delta_{n,p-1})\\
&-\frac{J}{\lambda_p}(\delta_{m-1,p}\delta_{n,p}+\delta_{m+1,p}\delta_{n,p})],
\end{align}
where $\mod(p,2)=0$.
We define new creation(annihilation) operators $\hat{d^\dagger}(\hat{d})$:
\begin{equation}
(\hat{d}_1, \hat{d}_2 \dots \hat{d}_L)^T \equiv \textbf{d}=e^{S} \textbf{c} . \label{c_to_d}
\end{equation}
We assume $\frac{J}{\lambda_p}$ is small enough in the large quasiperiodic potential strength region and $e^{-S}\approx I-S$.
By performing the S-W transformation and keeping the first order term, we can obtain the effective model for the quasiperiodic mosaic model:
\begin{equation}
\hat{H}= \textbf{c}^\dagger e^{-S} e^{S}H e^{-S} e^{S}\textbf{c}  =\textbf{d}^\dagger e^{S}H e^{-S} \textbf{d} \approx \hat{H}_e+\hat{H}_o,  \label{eff}
\end{equation}
where
\begin{equation*}
\hat{H}_{o}=\sum_{j=\text{odd}} - \frac{J^2}{\lambda_{j+1}} (\hat{d}_{j+2}^\dagger \hat{d}_{j}+\hat{d}_{j}^\dagger \hat{d}_{j+2} + \hat{d}^\dagger_j \hat{d}_j + \hat{d}^\dagger_{j+2} \hat{d}_{j+2})
\end{equation*}
and
\begin{equation*}
\hat{H}_{e}=\sum_{j=\text{even}}( \lambda_j + \frac{2 J^2}{\lambda_{j}} ) \hat{d}^\dagger_j \hat{d}_j +\frac{J^2(\lambda_j +\lambda_{j+2})}{2 \lambda_j \lambda_{j+2}} (\hat{d}_{j+2}^\dagger \hat{d}_{j}+\hat{d}_{j}^\dagger \hat{d}_{j+2}).
\end{equation*}
 The subscript of $e$ or $o$ indicates the Hamiltonian defined on the even or odd sites.
The details about S-W transformation can be found in Appendix \ref{SW}.
To the first order, $\hat{H}_e$ and $\hat{H}_o$ are decoupled and their dynamics seem to be independent of each other.
Next, we will study the properties the effective Hamiltonian and show how to use it to understand the special dynamical behaviors of quasiperiodic mosaic model.

\subsection{Properties of the effective Hamiltonian}
\begin{figure}[h]
    \centering
    \includegraphics[width=0.23\textwidth,height=0.2\textwidth]{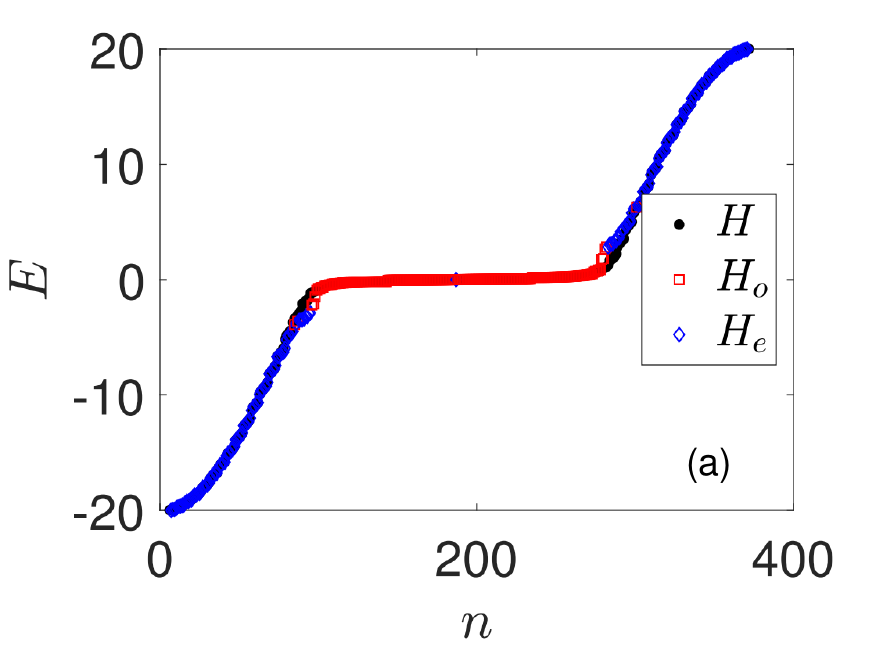}
    \includegraphics[width=0.24\textwidth,height=0.2\textwidth]{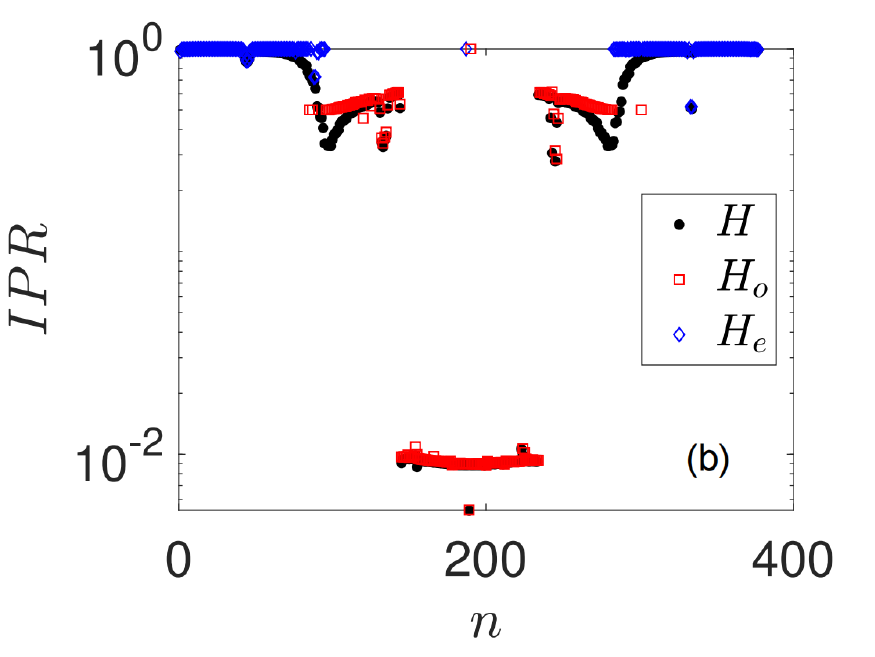}
    \caption{The comparison of eigenvalues (a) and eigenstates (b) obtained from the effective Hamiltonian and the quasiperiodic mosaic model. The black points, blue points and red points are the results from $\hat{H}$, $\hat{H}_o$ and $\hat{H}_e$ respectively. We choose $\lambda=10$ and $L=377$. $n$ labels the \textcolor{red}{$n$-{th}} eigenstate. }
    \label{compare1}
\end{figure}

First, we present the eigenstates and eigenvalues of the effective Hamiltonian, comparing them with the results obtained from exact diagonalization (ED). The results are illustrated in Fig. \ref{compare1}.
In Fig. \ref{compare1} (a), we depict the eigenvalues from the effective Hamiltonian and the quasiperiodic mosaic model, exhibiting  good agreement between them. It is evident that the eigenvalues of $\hat{H}_o$ contribute to the energy levels of the band center around $E=0$, while the eigenvalues of $\hat{H}_e$ contribute to the energy levels apart from the band center.
In Fig. \ref{compare1} (b), we show the IPR for the effective Hamiltonian and quasiperiodic mosaic model. The effective Hamiltonian can capture the extended-localized properties of a large number of states.
Although some eigenstates at the edge of the energy spectrum $\hat{H}_o$ deviate from the true eigenstates, we will show these eigenstates are localized and thus do not contribute to the dynamical behaviors.

From Eq.(\ref{eff}), we see that the effective Hamiltonian of the quasiperiodic mosaic model is divided into two decoupled chains:
(i) $\hat{H}_e$ distributes on the even sites and contains the onsite potential and hopping terms between even sites.
Since the energy scale of the onsite potential $O(\lambda)$ is much larger than the energy scale of the hopping term $O(\frac{1}{\lambda})$,
all the eigenstates of $\hat{H}_e$ are localized. 
\begin{figure}[h]
    \centering
    \includegraphics[width=0.46\textwidth,height=0.2\textwidth]{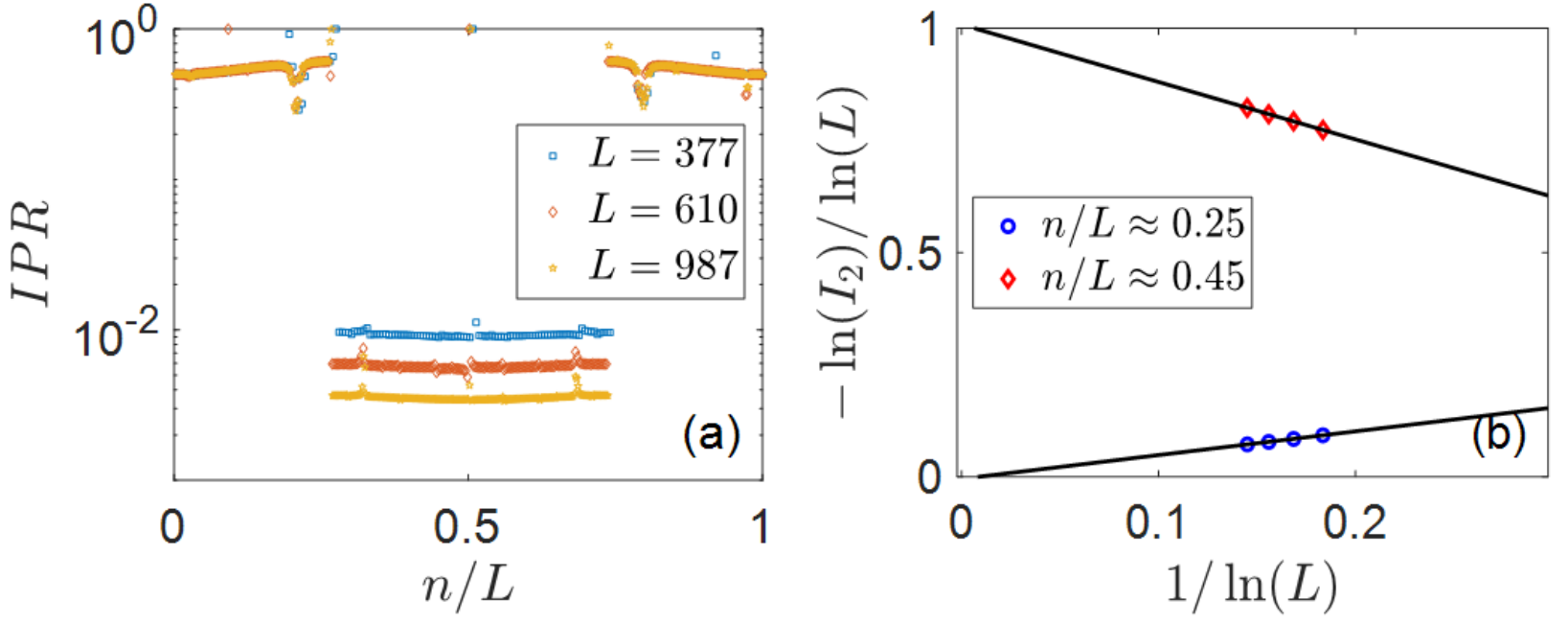}
    \caption{(a) IPR for eigenstates of $\hat{H}_o$ with  different system sizes. We choose $\lambda=10$ and $L=377,~610,~987$. (b) Finite-size scaling analysis for different eigenstates.}
    \label{scaling}
\end{figure}
(ii) $\hat{H}_o$ distributes on the odd sites.
Both the onsite potential and  hopping term have been modulated by the quasiperiodic potential equally.
As shown in Fig. \ref{scaling}(a): the IPRs for the eigenstates  of $\hat{H}_o$ in the middle of the energy spectrum is much smaller than that at the two sides of the spectrum, indicating that eigenstates in the middle and sides are extended and localized, respectively. To see it more clearly,
in Fig. \ref{scaling}(b), we perform finite-size scaling analysis on typical eigenstates in different regions using the moments $D_2$ (Eq.\ref{moment})
to characterize the distribution information of the eigenstate.
For our analysis, we take the logarithm of both sides of  Eq.(\ref{moment}) and obtain
\begin{align}
-\ln(I_2)/\ln(L)=-c/\ln(L)+D_2 .
\end{align}
We determine $D_2$ by conducting a linear fit to the curve in the two-dimensional space spanned by $-\frac{\ln(I_2)}{\ln(L)}$ and $-\frac{1}{\ln(L)}$. We find that the fractal dimension of the typical eigenstate (red diamond markers) in the middle of the spectrum tends to 1 as \( L \to \infty \), indicating that the eigenstate is extended. In contrast, the fractal dimension tends to 0 for the eigenstate (blue circle markers) at the edges of the spectrum, signifying that the eigenstate is localized. These results suggest the existence of mobility edges in the spectrum of \( \hat{H}_o \).

From the effective Hamiltonian, we gain additional insights:
(i) All the extended eigenstates can be traced back to the term $\hat{H}_o$. Despite the fact that the energy width of the extended states narrows as $\lambda$ increases, the proportion of extended states remains constant with the quasiperiodic potential strength.
(ii) Two types of localized eigenstates exist in the quasiperiodic mosaic model. They distribute on different sites and have different physical origins.
The first type of localized eigenstates originates from $\hat{H}_e$. These states are distributed at sites where the onsite potential is nonzero. Considering an eigenstate $|\psi_j\rangle$ with a localized center at the $j$-th site, its eigenvalue can be approximated as $\lambda_j+\frac{2J^2}{\lambda_j}$ which is proportional to quasiperiodic potential  strength $\lambda$ in the large quasiperiodic potential strength case.
The second type of localized eigenstates originates from $\hat{H}_o$. These eigenstates mainly distribute on sites where the potential is zero.  The eigenvalues of this eigenstate are inversely proportional to the quasiperiodic potential strength $\lambda$.

It should be noted that some onsite potentials may not be large enough due to the specific properties of the quasiperiodic potential. However, our numerical results indicate that, although in cases where the  S-W transformation may not be strictly applicable at a few sites, the effective Hamiltonian can still well capture the system's dynamical properties.
Further discussions on the effectiveness of effective Hamiltonian are provided in Appendix \ref{effectiveness}.

\subsection{Understanding the wave packet dynamics from the effective Hamiltonian}

The effective Hamiltonian Eq.(\ref{eff}) provides a clear physical picture to understand the wave packet dynamics of quasiperiodic mosaic model.
Since all eigenstates of $\hat{H}_e$ are localized,  $\hat{H}_e$ does not contribute to the expansion of the wave packet. In contrast, the eigenstates of $\hat{H}_o$ include both localized and extended eigenstates, the dynamics of $\hat{H}_o$ are similar to the general model with mobility edges: the wave packet will expand but part of the wave function remains at the initial position.
The wave packet at a given time $t$ can be written as:
\begin{align}
|\psi(t)\rangle= e^{-i \hat{H} t}|\psi(0) \rangle.
\end{align}
Replacing the Hamiltonian with the effective Hamiltonian and considering $\hat{H}_e$ and $\hat{H}_o$ are decoupled, we can discuss their dynamics separately:
\begin{align}
    e^{-i \hat{H} t}|\psi(0) \rangle & \approx  e^{-i (\hat{H}_{e}+\hat{H}_{o}) t} |\psi(0) \rangle \\
    & = e^{-i \hat{H}_{e} t}  (P_e |\psi(0) \rangle) +  e^{-i \hat{H}_{o} t}  (P_o |\psi(0) \rangle), \notag
\end{align}
where $P_o(P_e)$ represents the projection into the eigenstates of  $\hat{H}_o(\hat{H}_e)$.

Since  the wave packet dynamics is dominated by $\hat{H}_o$, the wave function only propagates along  the odd sites in the dressed basis ($d^\dagger_j|0\rangle$). For the initial state with the Gaussian wave packet located at the center of the lattice, the particle number distribution $n^d_j=  \langle \psi(t)|\hat{d}^{\dagger}_j \hat{d}_j |\psi(t) \rangle$ out of the initial wave packet only distributes on the odd sites. Nevertheless,
according to the S-W transformation Eq.(\ref{c_to_d}),  the creation (annihilation) operator on the even sites in the dressed basis can be related to the operators on both even and odd sites in the original basis.
Up to the second order, the density distribution on the even sites in the original basis can be explicitly expressed as
\begin{align}
   & \langle \psi(t)| \hat{c}_j^\dagger \hat{c}_j | \psi(t) \rangle \notag \\
   \approx &  \frac{J^2}{\lambda_j^2} \langle \psi(t)| \hat{d}_{j-1}^\dagger \hat{d}_{j-1} |\psi(t) \rangle
     + \frac{J^2}{\lambda_j^2} \langle \psi(t)| \hat{d}_{j+1}^\dagger \hat{d}_{j+1} | \psi(t) \rangle + \notag \\
    & \frac{J^2}{\lambda_{j}^2} (\langle \psi(t) | \hat{d}^\dagger_{j-1} \hat{d}_{j+1}| \psi(t) \rangle
     +\langle \psi(t)| \hat{d}^\dagger_{j+1} \hat{d}_{j-1}| \psi(t) \rangle),
\end{align}
whereas density distribution on the odd sites are given by
\begin{align}
    & \langle  \psi(t) |\hat{c}_j^\dagger \hat{c}_j| \psi(t) \rangle \notag\\
    \approx &(1-\frac{ J^2}{\lambda_{j+1}^2}- \frac{ J^2}{\lambda_{j-1}^2}) \langle \psi(t) |\hat{d}_{j}^\dagger \hat{d}_j | \psi(t) \rangle \notag\\
     & - \frac{J^2}{2\lambda_{j-1}^2} (\langle \psi(t)| \hat{d}^\dagger_{j-2} \hat{d}_{j}| \psi(t) \rangle +
     \langle \psi(t)| \hat{d}^\dagger_{j} \hat{d}_{j-2}| \psi(t) \rangle) \notag\\
     & - \frac{J^2}{2\lambda_{j+1}^2} (\langle \psi(t)| \hat{d}^\dagger_{j+2} \hat{d}_{j}| \psi(t) \rangle + \langle \psi(t)| \hat{d}^\dagger_{j} \hat{d}_{j+2}| \psi(t) \rangle) .
\end{align}
The details of this relationship can be found in Appendix.\ref{transform}.
The particle number distribution $n_j$ at even and odd sites all dates back to $n^d_j$ with $j=odd$.
Naturally, the wave function in odd and even sites will have the same propagation speed determined by $\hat{H}_{o}$. Since
 $n_j(t) \propto \frac{J^2}{\lambda^2} n^d_{j\pm1}(t)$ for $j=\text{even}$ and $n_j(t) \propto  n^d_{j}(t)$ for $j=\text{odd}$, the density distribution of odd and even sites have different orders of magnitude  and the ratio of even and odd density distribution is proportional to $\frac{J^2}{\lambda^2}$.

From the expression of $\hat{H}_o$, we observe that $\frac{1}{\lambda}$ is an overall factor of the Hamiltonian of $\hat{H}_o$.
It follows that the dynamics governed by $\hat{H}_o$ is invariant after a time scale $t \rightarrow t/\lambda$.
This is consistent with the numerical results as shown in the inset of Fig.2.
The wave packet dynamics governed by the effective Hamiltonian is also shown in Fig.\ref{figure1}(e) with the cyan and magenta lines, exhibiting  excellent agreement with the wave packet dynamics of the original Hamiltonian. Our results suggest that
the effective Hamiltonian can describe the dynamical behaviors well for the mosaic model in the large quasiperiodic potential strength region.

\section{Summary and discussion \label{E}}
In summary, we investigated the  dynamics of a quasiperiodic mosaic model and unravel its peculiar dynamical behaviors in the large quasiperiodic potential region,  exhibiting obvious different density distribution on even and odd sites. Our results also suggest that time scales of dynamical evolution  are determined by the strength of the quasiperiodic potential linearly.
Applying the S-W transformation, we derived an effective Hamiltonian in the limit of large quasiperiodic potential strength. The effective Hamiltonians can be divided into two decoupled Hamiltonians $\hat{H}_e$ and $\hat{H}_o$ defined on the even and odd sites, respectively. While all eigenstates of $H_e$ are localized, the eigenstates of $\hat{H}_o$ include both localized and extended eigenstates separated by mobility edge. The effective model describes the dynamical behaviors well for the mosaic model in the large quasiperiodic potential region. Our paper provides a clear picture for understanding the peculiar dynamical behaviors in the quasiperiodic mosaic lattice.

Although in this paper we only study the particular case of quasiperiodic mosaic models with $\kappa=2$ \cite{MosaicAA}, our scheme can directly apply to study dynamics in other cases with $\kappa>2$. Considering the case of $\kappa=3$, we can still apply S-W transformation to derive an effective Hamiltonian in the limit of large quasiperiodic potential strength, which is described by two weakly coupled chains (see Appendix.\ref{kappa3}). We anticipate that the wavepacket dynamics will also exhibit distinct density distributions on the modulated and unmodulated sites.

\acknowledgements
This work is supported by National Key Research and Development Program
of China (Grant No. 2021YFA1402104 and 2023YFA1406704), the NSFC under
Grants No. 12174436 and No. T2121001, and the Strategic Priority Research
Program of Chinese Academy of Sciences under Grant No. XDB33000000.

\appendix
\section{Two-point correlation function  \label{Imbalance}}
Here we give the details for the calculation of the time evolution of the correlation function following the method in Ref.\cite{calculate_Imbalance, calculate_Imbalance1,calculate_Imbalance2}.
First, we consider the time evolution of the correlation function $\langle \psi(t)| \hat{c}^\dagger_i \hat{c}_j| \psi(t) \rangle $:
\begin{align*}
    \langle \psi(t) | \hat{c}^\dagger_i \hat{c}_j |\psi(t) \rangle &=
    \langle \psi(0) |e^{i \hat{H} t} \hat{c}^\dagger_i \hat{c}_j e^{-i \hat{H} t}|\psi(0) \rangle  \\
    & = \langle e^{i \hat{H} t} \hat{c}^\dagger_i e^{-i \hat{H} t} e^{i \hat{H} t}  \hat{c}_j e^{-i \hat{H} t} \rangle,
\end{align*}
where $\langle \dots \rangle$ means $\langle \psi(0)|\dots|\psi(0) \rangle$.
Then we consider a transformation:
$$
\hat{c}_j=U_{jk}\hat{d}_k,
$$
$$
\hat{c}^\dagger_j=U^*_{jk}\hat{d}^\dagger_k,
$$
and the Hamiltonian can be written as
$$
\hat{H}=\sum_k\epsilon_k \hat{d}_k^\dagger \hat{d}_k,
$$
where $\epsilon_k$ are the eigenvalues of the Hamiltonian.
Then we can get the evolution of $\hat{d}_k$ and  $\hat{d}^\dagger_k$:
$$
e^{i \hat{H} t} \hat{d}_k^\dagger e^{-i \hat{H} t} =e^{i\epsilon_k t} \hat{d}_k^\dagger,
$$
$$
e^{i \hat{H} t} \hat{d}_k e^{-i \hat{H} t} =e^{- i\epsilon_k t} \hat{d}_k .
$$
The correlation function can be simplified as:
 \begin{align*}
    &\langle \psi(t) | \hat{c}^\dagger_i \hat{c}_j |\psi(t) \rangle \\
     =& \langle e^{i \hat{H} t} \hat{c}^\dagger_i e^{-i \hat{H} t} e^{i \hat{H} t}  \hat{c}_j e^{-i \hat{H} t} \rangle \\
    = &\sum_{k_1,k_2} \langle e^{i \hat{H} t} U^*_{i k_1} \hat{d}^\dagger_{k_1} e^{-i \hat{H} t} e^{i \hat{H} t}  U_{j k_2}\hat{d}_{k_2} e^{-i \hat{H} t} \rangle \\
    =& \sum_{k_1,k_2} U^*_{i k_1} U_{j k_2} \langle e^{i \hat{H} t}  \hat{d}^\dagger_{k_1} e^{-i \hat{H} t} e^{i \hat{H} t}  \hat{d}_{k_2} e^{-i \hat{H} t} \rangle \\
    =& \sum_{k_1,k_2} U^*_{i k_1} U_{j k_2} e^{i (\epsilon_{k_1}-\epsilon_{k_2}) t} \langle   \hat{d}^\dagger_{k_1}    \hat{d}_{k_2}  \rangle .
\end{align*}

Consider the inverse transformation between $\hat{d}(\hat{d}^\dagger)$ and $\hat{c}(\hat{c}^\dagger)$:
$$
\hat{d}_j=(U^{-1})_{jk}\hat{c}_k,
$$
$$
\hat{d}^\dagger_j=(U^{-1})^*_{jk}\hat{c}^\dagger_k,
$$
the correlation function can be written as
\begin{align*}
    &\langle \psi(t) | \hat{c}^\dagger_i \hat{c}_j |\psi(t) \rangle \\
    =& \sum_{k_1,k_2} U^*_{i k_1} U_{j k_2} e^{i (\epsilon_{k_1}-\epsilon_{k_2}) t} \langle   \hat{d}^\dagger_{k_1}    \hat{d}_{k_2}  \rangle \\
    =& \sum_{k_1,k_2,a,b} U^*_{i k_1} (U^{-1})^*_{k_1 a} U_{j k_2} (U^{-1})_{k_2 b} e^{i (\epsilon_{k_1}-\epsilon_{k_2}) t} \langle   \hat{c}^\dagger_{a}    \hat{c}_{b}  \rangle.
\end{align*}

Next we demonstrate how to calculate $\langle  \hat{c}^\dagger_a \hat{c}_b  \rangle$:
the initial state $|\psi(0) \rangle$ is chosen as:
$$
|\psi(0)\rangle= \prod_{\delta=1}^{L/2}\hat{c}^\dagger_{2\delta}|0\rangle,
$$
where $|0\rangle$is the vacuum and we rewrite it as:
$$
|\psi(0)\rangle= \prod_{\delta=1}^{L/2}\sum_{\sigma=1}^L P_{\sigma\delta} \hat{c}^\dagger_\sigma|0\rangle .
$$
To calculate the correlation function, we define two matrices $P^A$ and $P^B$. $P^A$ and $P^B$ are both $L \times (L/2+1) $ matrices and they satisfy:
$$
P^A_{\sigma \delta}= P_{\sigma \delta}, \quad \text{for} \quad \delta \leq L/2
$$
and
\begin{align*}
P^A_{\sigma L/2+1}= 1, \quad \text{for} \quad \sigma=a, \\
P^A_{\sigma L/2+1}= 0, \quad \text{for} \quad \sigma \neq a ,
\end{align*}
as well as
$$
P^B_{\sigma \delta}= P_{\sigma \delta}, \quad \text{for} \quad \delta \leq L/2
$$
and
\begin{align*}
P^B_{\sigma L/2+1}= 1, \quad \text{for} \quad \sigma=b, \\
P^B_{\sigma L/2+1}= 0, \quad \text{for} \quad \sigma \neq b .
\end{align*}

Then $\langle \hat{c}_b \hat{c}_a^\dagger \rangle$ can be gotten from
$$
\langle \hat{c}_b \hat{c}_a^\dagger \rangle=\det((P^A)^\dagger P^B),
$$
which gives rise to
$$
\langle \hat{c}_a^\dagger \hat{c}_b  \rangle=\delta_{ab} I-\det((P^A)^\dagger P^B).
$$
It follows the expression of the correlation function given by
\begin{align*}
      &\langle \psi(t) | \hat{c}^\dagger_i \hat{c}_j |\psi(t) \rangle \\
    = \sum_{k_1,k_2,a,b} & U^*_{i k_1} (U^{-1})^*_{k_1 a} U_{j k_2} (U^{-1})_{k_2 b} e^{i (\epsilon_{k_1}-\epsilon_{k_2}) t} \\& \{ \delta_{ab} I-\det[(P^A)^\dagger P^B]\}.
\end{align*}
We simplify this expression by writing it as the matrix form:
\begin{align*}
      &\langle \psi(t) | \hat{c}^\dagger_i \hat{c}_j |\psi(t) \rangle \\
    =&  U^* e^{i\Lambda t} (U^{-1})^* \mathcal{L} (U^{-1})^t e^{- i\Lambda t} U^T,
\end{align*}
where
$$
\mathcal{L}_{ab}=\delta_{ab} I-\det[(P^A)^\dagger P^B].
$$

\section{S-W transformation \label{SW}}
After choosing the new creation (annihilation) operator $\hat{d}_j^\dagger(\hat{d}_j)$,
the first quantized Hamiltonian can be written as:
$$
H'=e^S H e^{-S}.
$$
Here $S$ is a $L \times L$ anti-Hermitian matrix with the matrix element given by
\begin{align*}
S_{mn}=&\sum_p[\frac{J}{\lambda_p}(\delta_{m,p}\delta_{n,p+1}+\delta_{m,p}\delta_{n,p-1})\\
&-\frac{J}{\lambda_p}(\delta_{m-1,p}\delta_{n,p}+\delta_{m+1,p}\delta_{n,p})],
\end{align*}
where $\mod(p,2)=0$.
In the large quasiperiodic potential strength region, $\lambda_p \gg J$  and $J/\lambda_p$ can be regarded as a small quantity.
Utilizing the Baker-Campbell-Hausdorff formula, we can get the expression of the $H'$ approximately:
$$
H'=H+[S,H]+\frac{1}{2}[S,[S,H]]+\dots.
$$
We keep the lowest order term of $\frac{1}{\lambda}$ to get the effective Hamiltonian. The effective Hamiltonian can be written as:
\begin{equation*}
H_{\text{eff}}=H_e+H_o,
\end{equation*}
where
\begin{align*}
    (H_o)_{mn}=\sum_{j=\text{odd}} - \frac{J^2}{\lambda_{j+1}}  ( & \delta_{m,j}\delta_{n,j+2}  +\delta_{m,j+2} \delta_{n,j}\\
    &+ \delta_{m,j} \delta_{n,j}+\delta_{m,j+2}\delta_{n,j+2} )
\end{align*}
and
\begin{align*}
(H_e)_{mn}=& \sum_{j=\text{even}}[(\lambda_j + \frac{2 J^2}{\lambda_{j}} ) \delta_{m,j}\delta_{n,j} \\
& +\frac{J^2(\lambda_j +\lambda_{j+2})}{2 \lambda_j \lambda_{j+2}} (\delta_{m,j}\delta_{n,j+2}  +\delta_{m,j+2} \delta_{n,j})].
\end{align*}
Then we can write the Hamiltonian as
\begin{equation*}
\hat{H}_o=\sum_{j=\text{odd}} - \frac{J^2}{\lambda_{j+1}} (\hat{d}_{j+2}^\dagger \hat{d}_{j}+\hat{d}_{j}^\dagger \hat{d}_{j+2} + \hat{d}^\dagger_j \hat{d}_j + \hat{d}^\dagger_{j+2} \hat{d}_{j+2})
\end{equation*}
and
\begin{equation*}
\hat{H}_e=\sum_{j=\text{even}}( \lambda_j + \frac{2 J^2}{\lambda_{j}} ) \hat{d}^\dagger_j \hat{d}_j +\frac{J^2(\lambda_j +\lambda_{j+2})}{2 \lambda_j \lambda_{j+2}} (\hat{d}_{j+2}^\dagger \hat{d}_{j}+\hat{d}_{j}^\dagger \hat{d}_{j+2}).
\end{equation*}
We see that the Hamiltonian are divided into two decoupled chains. $\hat{H}_o$  includes both on-site quasiperiodic potential and off-diagonal quasiperiodic hopping terms.
The energy scale of diagonal term and
off-diagonal term are the same, proportional to $\frac{J^2}{\lambda}$.
For $\hat{H}_e$, it also includes on-site quasiperiodic potential and quasiperiodic hopping terms,
but with different energy scales.

\section{Effectiveness of unitary transformation \label{effectiveness}}
For quasiperiodic systems, even when $\lambda$ is large, there are still some sites whose potential strength is close to the hopping strength. This implies that the S-W transformation for these sites may not be ideal. The number of such sites increases linearly with the system size and decreases with $\lambda$.

\begin{figure}[h]
    \centering
    \includegraphics[width=0.23\textwidth,height=0.2\textwidth]{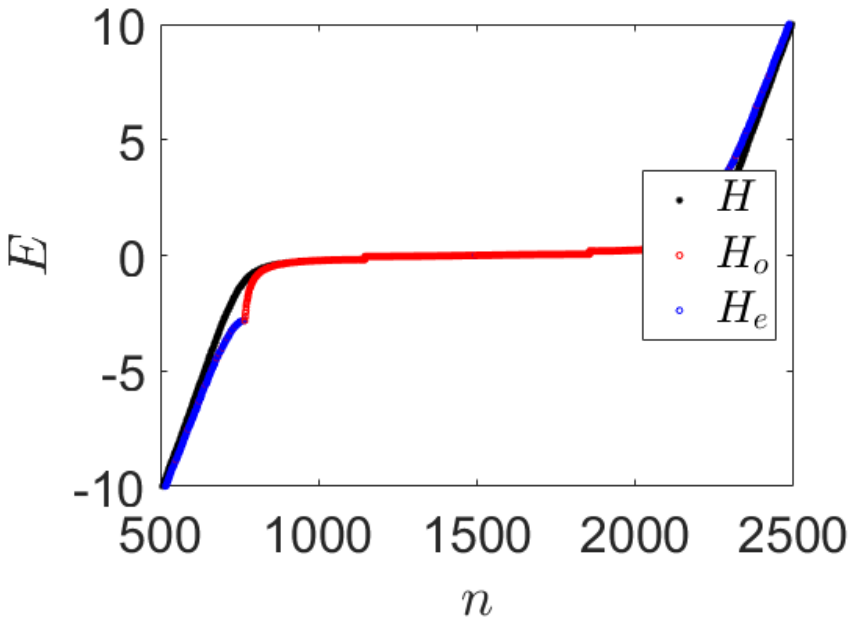}
    \includegraphics[width=0.24\textwidth,height=0.2\textwidth]{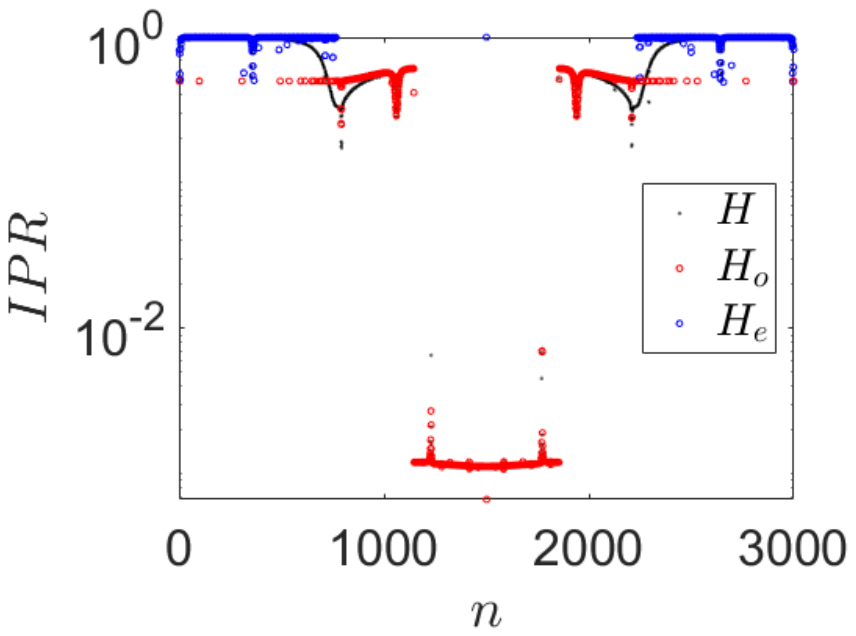}
    \caption{The comparison of eigenvalues(a) and eigenstates(b) obtained from the effective Hamiltonian and the quasiperiodic mosaic model. The black points, blue points and red points are the results from $\hat{H}$, $\hat{H}_e$ and $\hat{H}_o$ respectively. We choose $\lambda=10$ and $L=3\times10^3$. $n$ labels the \textcolor{red}{$n${th}}  eigenstate. }
    \label{compare2}
\end{figure}

In this paper, we investigate the impact of increasing the system size, denoted as $L$, on the effectiveness of the effective Hamiltonian under the same $\lambda$. Specifically, we choose $L=3\times10^3$ and compare the eigenvalues and IPR obtained from the effective Hamiltonian and exact ED. The main results are depicted in Fig.\ref{compare2}.

Remarkably, our findings indicate that the effective Hamiltonian remains effective in describing both the eigenvalues and the localization-extension properties of the system even as $L$ increases.

These sites with on-site potential strength close to the hopping strength only deviate the localized states from the ED results, but do not affect the properties of the extended state.  Therefore, our effective model can describe the expansion of the wave packet well.

\section{Transformation between different basis \label{transform}}
Representing $\langle \psi(t)| \hat{c}_j^\dagger \hat{c}_j | \psi(t) \rangle$ with $\langle \psi(t)| \hat{d}_j^\dagger \hat{d}_j  | \psi(t) \rangle$, we get
\begin{align*}
    \langle \psi(t) | \hat{c}^\dagger_j \hat{c}_j | \psi(t) \rangle &
    =\sum_{mk} \langle \psi(t) |\hat{d}^\dagger_m (e^{S})_{mj} (e^{-S})_{jk} \hat{d}_k  |\psi(t) \rangle \\
    & =(e^{S})_{mj} (e^{-S})_{jk} \langle \psi(t)| \hat{d}^\dagger_m \hat{d}_k  | \psi(t) \rangle \\
    & =(e^{S})_{mj} (e^{S})_{kj} \langle \psi(t)| \hat{d}^\dagger_m \hat{d}_k | \psi(t) \rangle
\end{align*}
Consider the large quasiperiodic potential strength limit and expand $e^{S}$ :
$$
e^{S}\approx1+S+\frac{1}{2}S^2.
$$
Keeping to the second order terms, we can get the expression  $\langle \psi(t)| \hat{c}_j^\dagger \hat{c}_j | \psi(t) \rangle$ for even and odd sites:
(1) for even sites:
\begin{align*}
   &  \langle \psi(t) | \hat{c}_j^\dagger \hat{c}_j | \psi(t) \rangle \\
    =&(1-\frac{2 J^2}{\lambda_j^2}) \langle \psi(t) |\hat{d}_j^\dagger \hat{d}_j |\psi(t) \rangle \\
    & +\frac{J^2}{\lambda_j^2} \langle \psi(t) |\hat{d}_{j-1}^\dagger \hat{d}_{j-1} |\psi(t) \rangle + \frac{J^2}{\lambda_j^2} \langle \psi(t)| \hat{d}_{j+1}^\dagger \hat{d}_{j+1}|\psi(t) \rangle \\
    &- \frac{J}{\lambda_j} (\langle \psi(t) |\hat{d}^\dagger_{j-1} \hat{d}_j | \psi(t)\rangle+ \langle \psi(t)| \hat{d}^\dagger_{j} \hat{d}_{j-1} |\psi(t)\rangle\\
    & -  \frac{J}{\lambda_j} (\langle \psi(t)| \hat{d}^\dagger_{j} \hat{d}_{j+1}|\psi(t) \rangle +\langle \psi(t)| \hat{d}^\dagger_{j+1} \hat{d}_{j} |\psi(t) \rangle )  \\
    &+ \frac{J^2}{\lambda_{j}^2} (\langle \psi(t) | \hat{d}^\dagger_{j-1} \hat{d}_{j+1} | \psi(t) \rangle +\langle \psi(t)| \hat{d}^\dagger_{j+1} \hat{d}_{j-1} | \psi(t) \rangle)\\
    &- \frac{J^2}{2\lambda_{j}\lambda_{j-2}} (\langle \psi(t)| \hat{d}^\dagger_{j} \hat{d}_{j-2} |\psi(t) \rangle +\langle \psi(t) |\hat{d}^\dagger_{j-2} \hat{d}_{j} | \psi(t) \rangle) \\
    &- \frac{J^2}{2\lambda_{j}\lambda_{j+2}} (\langle \psi(t) |\hat{d}^\dagger_{j} \hat{d}_{j+2} |\psi(t) \rangle +\langle \psi(t)| \hat{d}^\dagger_{j+2} \hat{d}_{j} |\psi(t) \rangle),
\end{align*}
(2) for odd sites:
\begin{align*}
    & \langle \psi(t) | \hat{c}_j^\dagger \hat{c}_j |\psi(t) \rangle \\
    &=(1-\frac{ J^2}{\lambda_{j+1}^2}- \frac{ J^2}{\lambda_{j-1}^2}) \langle \psi(t) |\hat{d}_{j}^\dagger \hat{d}_j |\psi(t) \rangle \\
    &+\frac{J^2}{\lambda_{j-1}^2} \langle \psi(t)| \hat{d}_{j-1}^\dagger \hat{d}_{j-1} |\psi(t) \rangle+ \frac{J^2}{\lambda_{j+1}^2} \langle \psi(t)| \hat{d}_{j+1}^\dagger \hat{d}_{j+1} |\psi(t)\rangle \\
    &- \frac{J}{\lambda_{j-1}} (\langle \psi(t)| \hat{d}^\dagger_{j-1} \hat{d}_{j}| \psi(t) \rangle+ \langle \psi(t)| \hat{d}^\dagger_{j} \hat{d}_{j-1}| \psi(t) \rangle)  \\
    & - \frac{J}{\lambda_{j+1}} (\langle \psi(t)| \hat{d}^\dagger_{j} \hat{d}_{j+1} |\psi(t) \rangle +\langle \psi(t)| \hat{d}^\dagger_{j+1} \hat{d}_{j} |\psi(t) \rangle )  \\
    &+ \frac{J^2}{\lambda_{j-1} \lambda_{j+1}} (\langle \psi(t)| \hat{d}^\dagger_{j-1} \hat{d}_{j+1}|\psi(t) \rangle +\langle \psi(t)| \hat{d}^\dagger_{j+1} \hat{d}_{j-1}|\psi(t) \rangle) \\
    & - \frac{J^2}{2\lambda_{j-1}^2} (\langle \psi(t)| \hat{d}^\dagger_{j-2} \hat{d}_{j} |\psi(t) \rangle +\langle \psi(t)| \hat{d}^\dagger_{j} \hat{d}_{j-2}| \psi(t) \rangle) \\
    & - \frac{J^2}{2\lambda_{j+1}^2} (\langle \psi(t)| \hat{d}^\dagger_{j+2} \hat{d}_{j} |\psi(t) \rangle +\langle \psi(t)| \hat{d}^\dagger_{j} \hat{d}_{j+2} |\psi(t) \rangle).
\end{align*}

Considering the wave packet only propagating in the odd sites in the effective Hamiltonian, we can reduce above expressions:
(1) for even sites:
\begin{align*}
    &\langle \psi(t) |\hat{c}_j^\dagger \hat{c}_j |\psi(t) \rangle \\
    & \approx  \frac{J^2}{\lambda_j^2} \langle \psi(t)| \hat{d}_{j-1}^\dagger \hat{d}_{j-1} |\psi(t)\rangle+ \frac{J^2}{\lambda_j^2} \langle \psi(t)| \hat{d}_{j+1}^\dagger \hat{d}_{j+1}| \psi(t) \rangle \\
    &+ \frac{J^2}{\lambda_{j}^2} (\langle \psi(t)| \hat{d}^\dagger_{j-1} \hat{d}_{j+1} |\psi(t) \rangle +\langle \psi(t)| \hat{d}^\dagger_{j+1} \hat{d}_{j-1}| \psi(t) \rangle),
\end{align*}
(2) for odd sites:
\begin{align*}
    &\langle \psi(t)| \hat{c}_j^\dagger \hat{c}_j |\psi(t) \rangle \\
    \approx &(1-\frac{ J^2}{\lambda_{j+1}^2}- \frac{ J^2}{\lambda_{j-1}^2}) \langle \psi(t)|\hat{d}_{j}^\dagger \hat{d}_j|\psi(t) \rangle \\
    & - \frac{J^2}{2\lambda_{j-1}^2} (\langle \psi(t)| \hat{d}^\dagger_{j-2} \hat{d}_{j}|\psi(t) \rangle +\langle \psi(t)| \hat{d}^\dagger_{j} \hat{d}_{j-2}|\psi(t) \rangle) \\
    & - \frac{J^2}{2\lambda_{j+1}^2} (\langle  \psi(t)|\hat{d}^\dagger_{j+2} \hat{d}_{j}|\psi(t) \rangle +\langle \psi(t)| \hat{d}^\dagger_{j} \hat{d}_{j+2}|\psi(t) \rangle).
\end{align*}
It follows that the density distribution of odd and even sites satisfy:
$$
\frac{n_{\text{odd}}(t)}{n_{\text{even}}(t)}\approx \frac{\lambda^2}{J^2}.
$$

\section{S-W transformation for $k=3$} \label{kappa3}

Here, we show that the S-W transformation can also be applied in the case of \(\kappa = 3\). We choose the system size to be an integer multiple of \(3\), and define a unit cell consisting of 3 sites, with the quasiperiodic potential placed on the second site. The Hamiltonian for the system is given by:
\[
\hat{H} = J \sum_j (\hat{c}_j^\dagger \hat{c}_{j+1} + \text{H.c.}) + 2 \sum_j \lambda_j \hat{n}_j,
\]
where \(\lambda_j\) is defined as:
\[
\lambda_j = 
\begin{cases}
\lambda \cos(2\pi \omega j + \phi), & \text{if } \mod(j+1, \kappa) = 0, \\
0, & \text{otherwise}.
\end{cases}
\]
with \(\kappa = 3\).
Following a similar procedure, we introduce an anti-Hermitian matrix:
\begin{align*}
S_{mn} = & \sum_p \frac{J}{\lambda_p} \left( \delta_{m,p} \delta_{n,p+1} + \delta_{m,p} \delta_{n,p-1} \right) \\ 
& - \frac{J}{\lambda_p} \left( \delta_{m-1,p} \delta_{n,p} + \delta_{m+1,p} \delta_{n,p} \right),
\end{align*}
where \(\mod(p+1, 3) = 0\). This transformation allows us to decompose the system into two weakly coupled chains, with the effective Hamiltonian expressed as:
\[
\hat{H} = \hat{H}_E + \hat{H}_L + \hat{H}_C,
\]
where the components are:
\[
\hat{H}_E = \sum_{j = 1, 4, 7, \dots}^{L-2}  - \frac{J^2}{\lambda_{j+1}} (\hat{d}_j^\dagger \hat{d}_j + \hat{d}_{j+2}^\dagger \hat{d}_{j+2} + \hat{d}_{j+2}^\dagger \hat{d}_j + \text{H.c.}) . 
\]
\[
\quad + \sum_{j = 1, 4, 7, \dots}^{L-5}  J (\hat{d}_{j+2}^\dagger \hat{d}_{j+3} + \text{H.c.}) ,
\]
\[
\hat{H}_L = \sum_{j = 2, 5, 8, \dots}^{L-1} \left( \lambda_j + \frac{2 J^2}{\lambda_j} \right) \hat{d}_j^\dagger \hat{d}_j,
\]
\[
\hat{H}_C = \sum_{j = 2, 5, 8, \dots}^{L-1} - \frac{8 J^3}{3 \lambda_j^2} (\hat{d}_{j+1}^\dagger \hat{d}_j + \hat{d}_{j-1}^\dagger \hat{d}_j + \text{H.c.}).
\]
The Hamiltonian \(\hat{H}_E\) includes weak disorder potentials and hopping terms. This part contributes to extended eigenstates, although the dynamics are more complex because the quasiperiodic potential strength is not a global factor. The term \(\hat{H}_L\) contributes to localized eigenstates, while \(\hat{H}_C\) represents weak coupling between the two chains. The overall dynamics of the system are determined by the interplay of these terms.


\end{document}